\documentclass[pra,twocolumn,floatfix,a4paper,superscriptaddress]{revtex4}

\usepackage{bm,color,amsmath,txfonts}
\usepackage{graphicx}
\usepackage{siunitx}
\usepackage{subfigure}
\usepackage{verbatim}
\usepackage{dcolumn}
\usepackage{bm}
\usepackage{epsf}
\usepackage{xcolor}
\usepackage{hyperref}
\usepackage{hhline}
\usepackage{float}
\usepackage{enumerate}
\usepackage{bbm}
\usepackage{lipsum}
\usepackage{mathrsfs}

\begin{document}
\title[Short Title]{Magnon squeezing by two-tone driving of a qubit in cavity-magnon-qubit systems}
\author{Qi Guo\footnote{E-mail: qguo@sxu.edu.cn}}
\affiliation{Zhejiang Province Key Laboratory of Quantum Technology and Device, School of Physics and State Key Laboratory of Modern Optical Instrumentation, Zhejiang University, Hangzhou 310027, China}
\affiliation{State Key Laboratory of Quantum Optics and Quantum Optics Devices, and College of Physics and Electronic Engineering, Shanxi University, Taiyuan, Shanxi 030006, People's Republic of China}
\author{Jiong Cheng}
\affiliation{Zhejiang Province Key Laboratory of Quantum Technology and Device, School of Physics and State Key Laboratory of Modern Optical Instrumentation, Zhejiang University, Hangzhou 310027, China}
\affiliation{Department of Physics, School of Physical Science and Technology, Ningbo University, Ningbo, 315211, China}
\author{Huatang Tan\footnote{E-mail: tht@mail.ccnu.edu.cn}}
\affiliation{Department of Physics, Huazhong Normal University, Wuhan 430079, China}
\author{Jie Li\footnote{E-mail: jieli007@zju.edu.cn}}
\affiliation{Zhejiang Province Key Laboratory of Quantum Technology and Device, School of Physics and State Key Laboratory of Modern Optical Instrumentation, Zhejiang University, Hangzhou 310027, China}

\begin{abstract}
We propose a scheme for preparing magnon squeezed states in a hybrid cavity-magnon-qubit system. The system consists of a microwave cavity that simultaneously couples to a magnon mode of a macroscopic yttrium-iron-garnet (YIG) sphere via the magnetic-dipole interaction and to a transmon-type superconducting qubit via the electric-dipole interaction. By far detuning from the magnon-qubit system, the microwave cavity is adiabatically eliminated. The magnon mode and the qubit then get effectively coupled via the mediation of virtual photons of the microwave cavity.     We show that by driving the qubit with two microwave fields and by appropriately choosing the drive frequencies and strengths, magnonic parametric amplification can be realized, which leads to magnon quadrature squeezing with the noise below vacuum fluctuation.  We provide optimal conditions for achieving magnon squeezing, and moderate squeezing can be obtained using currently available parameters. The generated squeezed states are of a magnon mode involving more than $10^{18}$ spins and thus macroscopic quantum states. The work may find promising applications in quantum information processing and high-precision measurements based on magnons and in the study of macroscopic quantum states.
\end{abstract}
\maketitle

\section{Introduction}

With the increasing improvement of experimental technology, the study of macroscopic quantum states has been attracting more and more attention since the Schr\"{o}dinger's cat state was proposed \cite{1}. In particular, cavity optomechanics (COM), exploring the interaction between electromagnetic fields and mechanical motion via radiation pressure, provides an ideal platform to prepare macroscopic quantum states~\cite{MA}. In the past decade, significant progress has been made in the field of COM in generating macroscopic quantum states of massive mechanical oscillators. These include the realization of the entangled states of a mechanical oscillator and an electromagnetic field~\cite{Palomaki}, the entangled states of two mechanical oscillators~\cite{mechanical1,mechanical2,mechanical3}, the squeezed states~\cite{qusqz} and superposition states~\cite{Wal,Fia} of mechanical motion, etc. In addition, nonclassical states, e.g., superposition states~\cite{Cleland}, Fock states~\cite{Chu}, cat states~\cite{Fadel1} and entangled states~\cite{ASN,Fadel2}, of macroscopic mechanical resonators can also be generated by coupling to and controlling the superconducting qubit.

In recent years, hybrid systems based on collective spin excitations (magnons) in macroscopic ferromagnetic crystals, such as yttrium-iron-garnet (YIG), have become a new platform to explore macroscopic quantum phenomena and develop novel quantum technologies \cite{2,Yuan,YLi}.  It was first proposed in cavity magnomechanics~\cite{16Tang,21,PRX21,Shen22} that macroscopic entangled states of magnons, photons and phonons can be created exploiting the dispersive magnetostrictive interaction~\cite{21}. Such nonlinear magnomechanical coupling can also be used to entangle two magnon modes~\cite{50}, two mechanical modes~\cite{HQian}, and generate squeezed states of magnons and phonons~\cite{33}.  It can also be exploited to achieve Einstein-Podolsky-Rosen steering between magnons, photons and phonons~\cite{44,45}, and quantum ground states of mechanical vibration~\cite{Ding,Jing21,FR}.
Apart from utilizing the nonlinear magnetostriction, many other mechanisms have been put forward in cavity magnonics to prepare macroscopic quantum states. Specifically, the nonlinear magnon-photon interaction in cavity optomagnonics is exploited to cool magnons~\cite{Bauer18}, and prepare magnon Fock~\cite{Fock19}, cat~\cite{34} and path-entangled~\cite{42} states, as well as the entangled states of magnons and optical photons~\cite{PRXQ,Xie}.   Dissipative coupling between magnons and microwave photons is used to generate a magnon-photon Bell state~\cite{Yuan20}. 
Anisotropy, together with conditional measurements on microwave cavity photons, is utilized to prepare a magnon cat state~\cite{SVK21}.   Kerr-type nonlinearities are adopted to entangle two magnon modes~\cite{GSA19,Yipu} and achieve one-way quantum steering between ferrimagnetic microspheres~\cite{Tan22}. Another approach is to use external quantum drives, e.g., single-mode or two-mode squeezed vacuum fields, which are employed to entangle two magnon modes~\cite{Jaya,Yu} and mechanical modes~\cite{JieQST}, and control one-way quantum steering~\cite{Yang21,Zhong21,Zhang22}.

The effective coupling of magnons with superconducting qubits via the mediation of microwave cavity photons can also provide necessary nonlinearity to prepare quantum states of magnons~\cite{2,22,Yuan}.
Due to the high controllability and scalability of the superconducting circuits, the study on the hybrid cavity-magnon-superconducting-qubit system has been receiving increasing attention in recent years.  Significant experimental progress has been made in this system. Specifically, strong coupling between a magnon and a superconducting qubit and magnon-vacuum induced Rabi splitting were demonstrated \cite{23}. Shortly afterwards, the quanta of a magnon mode in a millimeter-sized YIG sphere were resolved by using the magnon-qubit strong dispersive interaction \cite{25}. Working in the same dispersive regime, high-sensitivity detection of a single magnon in a YIG sphere with quantum efficiency of up to 0.71 was realized  \cite{26}.
Very recently, the superposition state of a single magnon and vacuum was deterministically generated \cite{XuDa}.  These successful experimental demonstrations have further stimulated the study on the quantum states in such a hybrid system. A series of theoretical proposals have been provided to explore quantum effects in the system, such as magnon blockade \cite{52,53,ZhouL,54,55,56}, continuous-variable \cite{43,49} and discrete-variable \cite{57,58,59,60,Zhang23} magnon entanglement and steering, magnon cat states \cite{35,36}, and so on. All of these indicate that the magnon-qubit system is a promising system to prepare various magnonic quantum states via manipulating the qubit.

Here, we show how to generate magnon squeezed states in such a cavity-magnon-qubit system. 
To date, only a few protocols have been offered in cavity magnonics to prepare magnon squeezed states. They can be achieved by exploiting the anisotropy or nonlinearities of the ferromagnet~\cite{Kamra,Mehrdad}, the mechanism of the ponderomotive-like squeezing~\cite{NSR}, the reservoir-engineering technique~\cite{76}, or the squeezed external drive fields~\cite{33,HYYuan21}.
Our approach differs from all the above mechanisms and is realized via two-tone driving of the superconducting qubit. It is akin to that used to produce squeezed light by two-tone driving of an atom~\cite{69}. The system is operating in the regime where the microwave cavity is far detuned from the magnon-qubit system and can thus be adiabatically eliminated. The qubit is simultaneously driven by two microwave fields. We show that by properly choosing the drive frequencies and strengths, the effective parametric amplification Hamiltonian can be obtained for the magnon mode, which leads to a two-magnon process and thus the squeezing of the magnon mode.  

The paper is organized as follows. In Sec.~\ref{model}, we describe the system and derive the effective Hamiltonian for the magnon mode, which gives rise to magnon quadrature squeezing. In Sec.~\ref{result}, we present the numerical results of the magnon squeezing, check the validity of our derived approximate model, provide the optimal drive conditions, and analyze the dissipation and thermal noise effects on the squeezing.  Lastly, we draw the conclusions in Sec.~\ref{conc}.

\begin{figure}[t]
\centering
\includegraphics[width=0.98\linewidth]{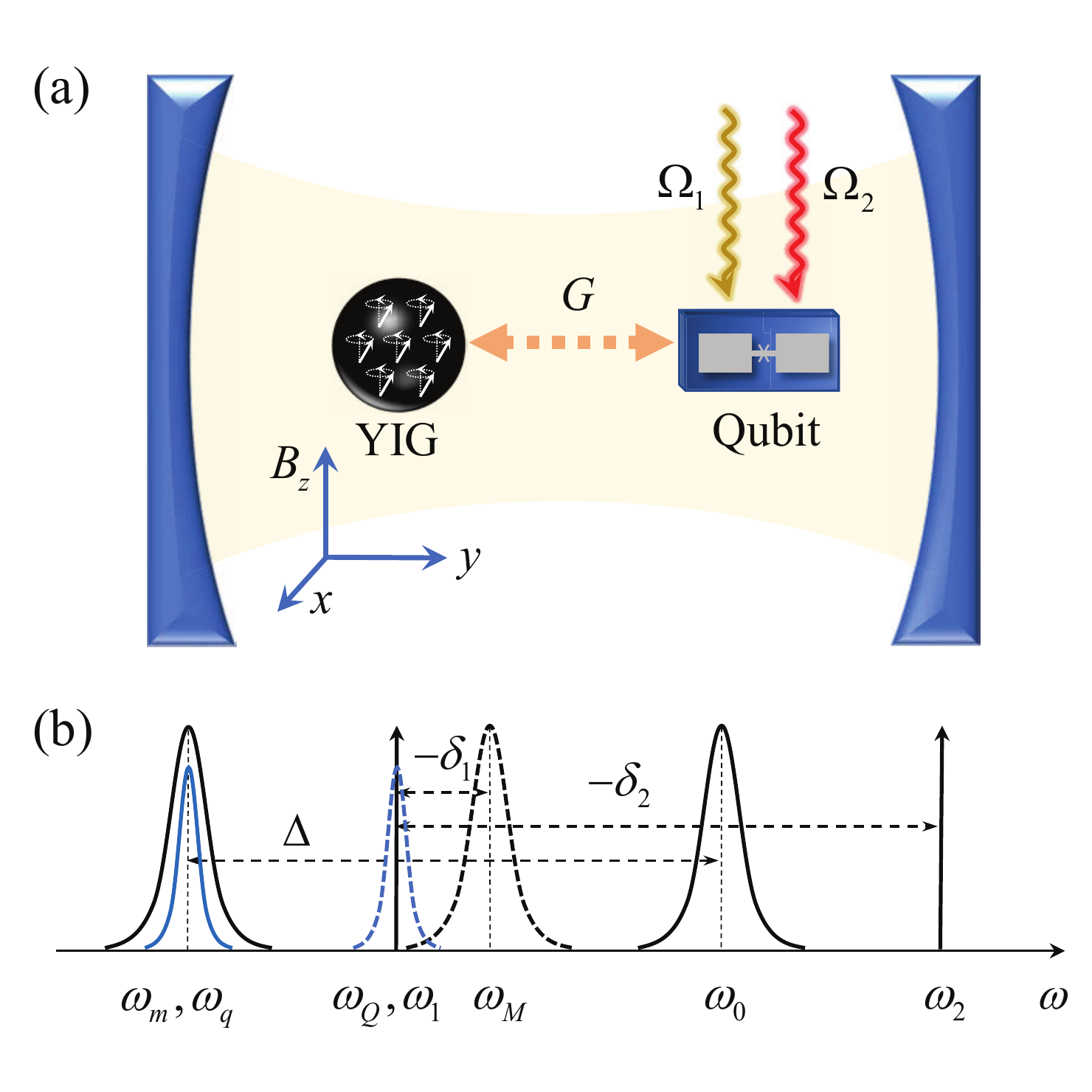}
\caption{(a) Schematic of the cavity-magnon-superconducting-qubit system. A microwave cavity couples to both a magnon mode of a macroscopic YIG sphere, which is placed in a uniform bias magnetic field $B_z$ ($z$ direction), and a superconducting qubit, which is driven by two microwave fields.  The magnon mode and the qubit get effectively coupled via the mediation of the microwave cavity. (b) Frequency spectrum of the system. The cavity with frequency $\omega_{0}$ is far-detuned from the magnon mode ($\omega_{m}$) and the qubit ($\omega_{q}$). The effective qubit transition frequency $\omega_{Q}$ is resonant with the drive field at frequency $\omega_{1}$, but is detuned by $\delta_{1}$ and $\delta_{2}$, respectively, from the effective magnon frequency $\omega_{M}$ and the drive field at frequency $\omega_{2}$.}
\label{f1}
\end{figure}

\section{The system and effective Hamiltonian}
\label{model}

The hybrid cavity-magnon-superconducting-qubit system, as depicted in Fig.~\ref{f1}(a), consists of a YIG sphere (e.g., with the diameter of 1 mm~\cite{XuDa}) and a transmon-type superconducting qubit that are placed inside a microwave cavity. The YIG sphere supports a magnon mode (collective motion of a large number of spins), which couples to the microwave cavity via the magnetic-dipole interaction and the latter further couples to the qubit via the electric-dipole interaction. The Hamiltonian of this tripartite system reads ($\hbar=1$)
\begin{eqnarray}\label{e1}
H&\!=&\!\omega_{0}a^{\dag}a+\frac{1}{2}\omega_{q}\sigma_{z}+\omega_{m}m^{\dag}m\cr
&&+g_{1}\left(a\sigma^{+}+a^{\dag}\sigma^{-}\right)+g_{2} \left(am^{\dag}+a^{\dag}m \right),
\end{eqnarray}
where $a$ ($a^{\dag}$) and $m$ ($m^{\dag}$) are the annihilation (creation) operators of the microwave cavity and the magnon mode, respectively, and $\omega_{0}$ and $\omega_{m}$ are their resonance frequencies.  We limit the subspace of the transmon-type qubit to the ground state $|g\rangle$ and the first-excited state $|e\rangle$, and the Pauli matrix $\sigma_{z}=|e\rangle\langle e|-|g\rangle\langle g|$, and $\sigma^{-}=|g\rangle\langle e|$ and $\sigma^{+}=|e\rangle\langle g|$ are the ladder operators of the qubit with transition frequency $\omega_{q}$.  The coupling strengths $g_{1}$ and $g_{2}$ are of the cavity-qubit and cavity-magnon systems, respectively.

For simplicity, we consider the situation where the qubit and the magnon are resonant, $\omega_{q}=\omega_{m}\equiv\omega$, and far-detuned from the microwave cavity, i.e., $\Delta=\omega_{0}-\omega \gg g_{1}, g_{2}$. This allows us to adiabatically eliminate the cavity mode and obtain the effective Jaynes-Cummings-type Hamiltonian of the magnon-qubit system \cite{2}, which is given by
\begin{eqnarray}\label{e2}
H_{\mathrm{eff}}=\frac{1}{2}\omega_{Q}\sigma_{z}+\omega_{M}m^{\dag}m+G\left(\sigma^{+}m+\sigma^{-}m^{\dag}\right),
\end{eqnarray}
where $\omega_{Q}=\omega+\frac{g_{1}^{2}}{\Delta}$ and $\omega_{M}=\omega+\frac{g_{2}^{2}}{\Delta}$ correspond to the effective frequencies of the qubit and the magnon mode, respectively (c.f. Fig.~\ref{f1}(b)), and $G=\frac{g_{1}g_{2}}{\Delta}$ denotes the effective magnon-qubit coupling. Such an effective Hamiltonian has been adopted in the experiments \cite{22,23,25,26,XuDa}.

We then apply two microwave fields to drive the qubit and the drive frequencies are $\omega_{1}=\omega_{Q}$ and $\omega_{2}$, and the corresponding driving strengths are $\Omega_{1}$ and $\Omega_{2}$. The Hamiltonian, in the interaction picture with respect to $\omega_{1} \big(\frac{1}{2} \sigma_{z}+m^{\dag}m \big) $, can be written as
\begin{eqnarray}\label{e3}
H_{1}= {-}\delta_{1}m^{\dag}m+\left(G\sigma^{+}m + \Omega_{1}\sigma^{+} + \Omega_{2}e^{i\delta_{2}t}\sigma^{+}+{\rm H.c.} \right),
\end{eqnarray}
where $\delta_{1}=\omega_{1}-\omega_{M}$ and $\delta_{2}=\omega_{1}-\omega_{2}$. Without loss of generality, $\Omega_{1}$ and $\Omega_{2}$ are assumed to be real.  To express the physics more straightforwardly, we adopt the qubit representation dressed by the drive field (of frequency $\omega_{1}$). By diagonalizing the driving Hamiltonian $V_{1}=\Omega_{1}(\sigma^{+}+\sigma^{-})$, the dressed states are expressed as
\begin{eqnarray}\label{e4}
|+\rangle=\frac{1}{\sqrt{2}} \left(|e\rangle+|g\rangle \right),\cr\cr
|-\rangle=\frac{1}{\sqrt{2}} \left(|e\rangle-|g\rangle \right).
\end{eqnarray}
Rewriting the Hamiltonian $H_{1}$ in terms of the dressed states, we obtain
\begin{eqnarray}\label{e5}
H_{2}\!\!&=&\!\! -\delta_{1}m^{\dag}m+\Omega_{1} \left(\sigma_{++}-\sigma_{--} \right)\cr
&&\!\!+\frac{1}{2} \Big[ \left(Gm+\Omega_{2}e^{i\delta_{2}t} \right) \left(\sigma_{++}{-}\sigma_{+-}{+}\sigma_{-+}{-}\sigma_{--} \right)\cr
&&\!\!+\left(Gm^{\dag}+\Omega_{2}e^{-i\delta_{2}t} \right) \left(\sigma_{++}{-}\sigma_{-+}{+}\sigma_{+-}{-}\sigma_{--} \right) \Big],
\end{eqnarray}
where we define $\sigma_{jk}=|j\rangle\langle k|$ ($j,k=+,-$). Working in the interaction picture with respect to $-\delta_{1}m^{\dag}m+\Omega_{1} \left(\sigma_{++}-\sigma_{--} \right)$ and taking $\Omega_{1}=-\frac{1}{2}\delta_{2}$, the Hamiltonian becomes
\begin{eqnarray}\label{e51}
H_{3}\!\!&=&\!\!\frac{1}{2}Gm\left[\left(\sigma_{++} {-} \sigma_{--} \right)e^{i\delta_{1}t}-\sigma_{+-}e^{-i(\delta_{2}-\delta_{1})t}+ \sigma_{-+}e^{i(\delta_{2}+\delta_{1})t} \right]\cr
&&\!\!+\frac{1}{2}\Omega_{2}\left[\left(\sigma_{++} {-} \sigma_{--} \right)e^{i\delta_{2}t}-\sigma_{+-}+\sigma_{-+}e^{i2\delta_{2}t} \right]+{\rm H.c.}.
\end{eqnarray}
Under the conditions of $|\delta_{2}|\gg \frac{G}{2}, \frac{\Omega_{2}}{2}, |\delta_{1}|$, we can take the rotating-wave approximation and obtain the following Hamiltonian
\begin{eqnarray}\label{e6}
H_{4}\,{=}\frac{1}{2}G \left(me^{i\delta_{1}t} {+} m^{\dag}e^{-i\delta_{1}t} \right) \left(\sigma_{++} {-} \sigma_{--} \right) {-} \frac{1}{2}\Omega_{2} \left(\sigma_{+-} {+} \sigma_{-+} \right) .
\end{eqnarray}
The second term $V_{2}=-\frac{1}{2}\Omega_{2}(\sigma_{+-}+\sigma_{-+})$ corresponds to the driving Hamiltonian associated with the second drive for the qubit. By diagonalizing $V_{2}$, we find that its eigenstates $(|+\rangle\pm|-\rangle)/\sqrt{2}$ are exactly the bare qubit states $|e\rangle$ and $|g\rangle$. Therefore, the Hamiltonian \eqref{e6} can be expressed in the initial qubit-state basis $\{|e\rangle, |g\rangle\}$ as
\begin{eqnarray}\label{e7}
H_{5}=-\frac{1}{2}\Omega_{2}\sigma_{z}+\frac{1}{2}G \left(me^{i\delta_{1}t}+m^{\dagger}e^{-i\delta_{1}t} \right)(\sigma^{+}+\sigma^{-}).
\end{eqnarray}
The Hamiltonian above in the interaction picture with respect to $-\frac{1}{2}\Omega_{2}\sigma_{z}$ then becomes
\begin{eqnarray}\label{e71}
H_{6}=\frac{1}{2}G m\left[\sigma^{+}e^{i(\delta_{1}-\Omega_{2})t}+\sigma^{-}e^{i(\delta_{1}+\Omega_{2})t}\right]+{\rm H.c.}.
\end{eqnarray}
Note that for $|\delta_{1}|\ll\Omega_{2}$, $\delta_{1}-\Omega_{2}<0$ and $\delta_{1}+\Omega_{2}>0$. According to the effective Hamiltonian theory \cite{78}, when the condition $|\delta_{1}\pm\Omega_{2}|\gg\frac{G}{2}$ is satisfied, the effective Hamiltonian is given by
\begin{eqnarray}\label{e72}
H_{\mathrm{eff}}=-iH_{6}(t)\int H_{6}(t')dt'.
\end{eqnarray}
Substituting Eq.~\eqref{e71} into Eq.~\eqref{e72} and ignoring the fast oscillation terms, we can obtain the following effective Hamiltonian
\begin{eqnarray}\label{e8}
H_{7}\!\!&=&\!\!\! \frac{G^2}{4} \Bigg[ \frac{1}{\delta_{1}{-}\Omega_{2}} \left(m^{\dagger}m\sigma_{z}{+}\sigma^{+}\sigma^{-} \right){+}\frac{1}{\delta_{1}{+}\Omega_{2}} \left(-m^{\dagger}m\sigma_{z}{+}\sigma^{-}\sigma^{+} \right)\cr\cr
&&\,\,\,\,\,\,+\frac{1}{\Omega_{2}}m^{2}\sigma_{z}e^{i2\delta_{1}t}+\frac{1}{\Omega_{2}}m^{\dagger 2}\sigma_{z}e^{-i2\delta_{1}t}  \Bigg].
\end{eqnarray}
For the case of the qubit being initially prepared in the state $|e\rangle$ (similarly, for the ground state $|g\rangle$), we obtain the parametric amplification Hamiltonian for the magnon mode in the interaction picture, i.e.,
\begin{eqnarray}\label{e9}
H_{8}=\chi\left[m^{2}e^{i \left(2\delta_{1}-\frac{\Omega_{2}G^{2}}{(\delta_{1}^{2}-\Omega_{2}^{2})} \right)t}+m^{\dagger 2}e^{-i \left(2\delta_{1}-\frac{\Omega_{2}G^{2}}{(\delta_{1}^{2}-\Omega_{2}^{2})} \right)t}\right],
\end{eqnarray}
where $\chi=G^2/(4\Omega_{2})$. This Hamiltonian describes a two-magnon process and can generate a magnon squeezed vacuum state. The squeezing direction in the phase space rotates due to the time dependence of the Hamiltonian. By appropriately choosing the parameters to have $\delta_{1}=\frac{\Omega_{2}G^{2}}{2(\delta_{1}^{2}-\Omega_{2}^{2})}$, i.e., $2\Delta\Omega_{2} {=}- g_{1}^{2}g_{2}^{2}/(g_{1}^{2}{-}g_{2}^{2})$, the Hamiltonian \eqref{e9} can be time-independent, which yields the normal parametric amplification Hamiltonian of $ \chi \big(m^{2}+m^{\dagger 2} \big)$.

\section{Results of magnon quadrature squeezing}
\label{result}

In Sec.~\ref{model}, we prove analytically that our mechanism can generate squeezing of the magnon mode and the derivation is performed without considering any dissipation of the system. In this section, we present the numerical results of the magnon squeezing by including dissipations of the system and using experimentally feasible parameters. We calculate the magnon squeezing by using the effective Hamiltonian \eqref{e9}, and compare it with that obtained using the original (full) Hamiltonian~\eqref{e3}, where no approximation is made. This allows us to check the validity of our model and determine the parameter regime where the effective Hamiltonian is a good approximation.

The squeezing denotes that the variance of the general quadrature of the magnon mode, $X=\cos \theta X_1 + \sin \theta X_2$, is below that of the vacuum noise, where $X_{1}=(m+m^{\dagger})/\sqrt{2}$ and $X_{2}=i(m^{\dagger}-m)/\sqrt{2}$ are the magnon amplitude and phase quadratures. In fact, the minimum variance of the quadrature $X$, i.e., $V_{\rm min}(X)$, can be obtained analytically using the time-independent parametric amplification Hamiltonian \eqref{e9} under precisely chosen parameters. Here, to be generic, we calculate the variance using the time-dependent Hamiltonian \eqref{e9}. 
The time dependence leads to the time-dependent optimal squeezing angle $\theta_{\rm opt}$, corresponding to the minimum variance and thus the maximum squeezing. However, $V_{\rm min}(X)$ can still be achieved by computing the minimum eigenvalue of the covariance matrix (CM) $\sigma$ of the two magnon quadratures $X_{1,2}$, i.e.,
\begin{equation}
V_{\rm min}(X)=\mathrm{min}\left\{\mathrm{eig}[\sigma] \right\}.
\end{equation}
The CM $\sigma$ is defined as
\begin{eqnarray}\label{e10}
\sigma=\left(\begin{array}{cc}
\sigma_{11}&\sigma_{12}\\
\sigma_{21}&\sigma_{22}\\
\end{array}
\right),
\end{eqnarray}
where $\sigma_{jk}=\mathrm{Tr}[\rho(X_{j}X_{k}+X_{k}X_{j})/2]-\mathrm{Tr}[\rho X_{j}]\mathrm{Tr}[\rho X_{k}]$ ($j,k=1,2$), and $\rho \equiv \rho_{}(t)$ is the density matrix of the system at time $t$. The optimal squeezing angle $\theta_{\rm opt}$ can be obtained from the CM $\sigma$, which is $\theta_{\rm opt} = \frac{1}{2} \arctan \frac{2 \sigma_{12}}{\sigma_{11}-\sigma_{22}} -\frac{\pi}{2}$.

\begin{figure}[t]
\centering
\hskip-0.3cm\includegraphics[width=\linewidth]{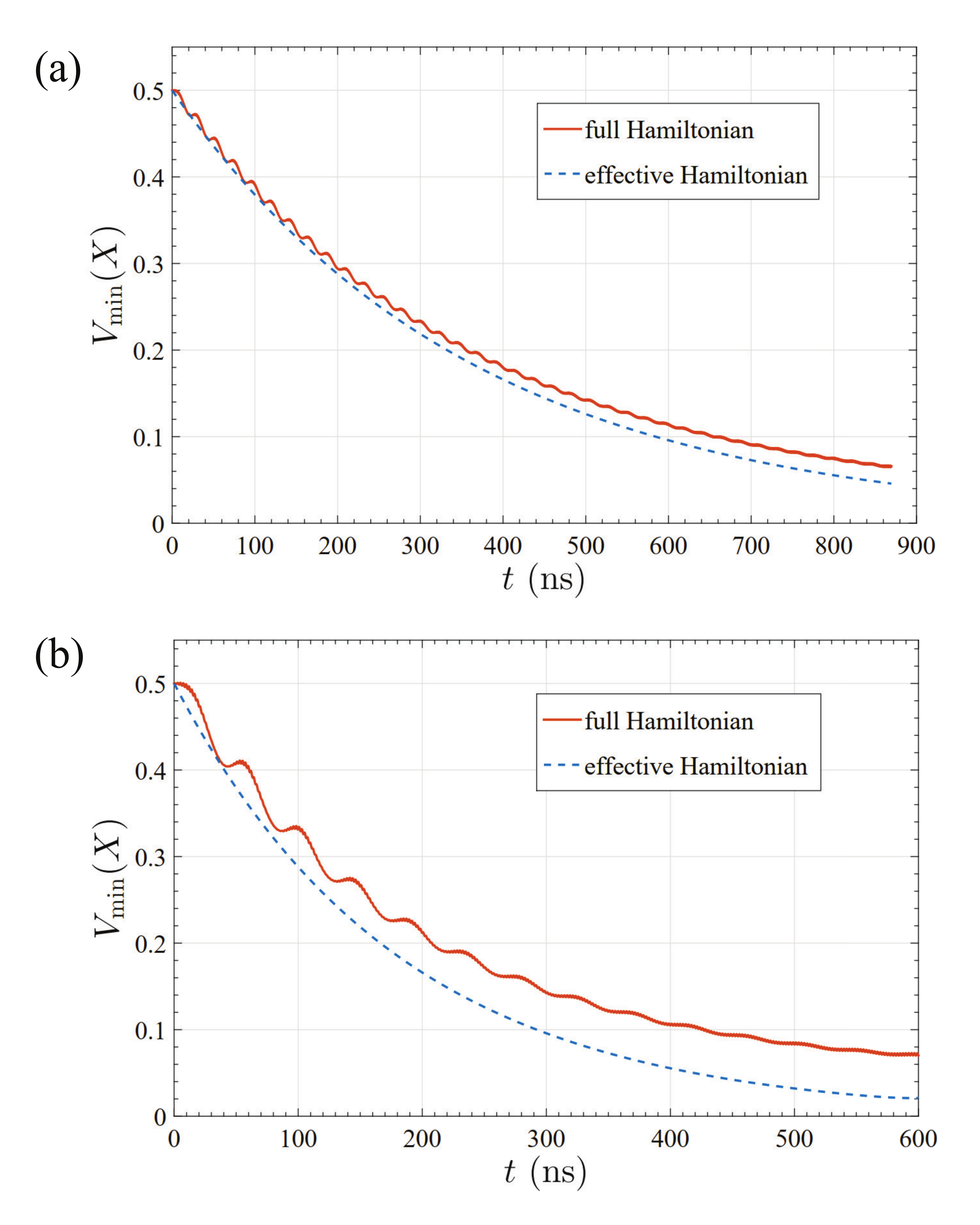}
\caption{Minimum variance $V_{\rm min}(X)$ of the magnon quadrature as a function of time $t$ for (a) $\Omega_{1}=10 \Omega_{2} =10^2G$; (b) $\Omega_{1}=10 \Omega_{2}=50G$. See text for the other parameters.}
\label{fig2}
\end{figure}

In Fig.~\ref{fig2}(a), we plot the minimum variance $V_{\rm min}(X)$ as a function of time $t$, where the solid (dashed) line corresponds to the result obtained using the full (effective) Hamiltonian~\eqref{e3} (\eqref{e9}).  We use experimentally feasible parameters \cite{23,25,26,XuDa}: $\omega_{0}/2\pi=7.5$ GHz, $\omega/2\pi=7.2$ GHz, $g_{1}/2\pi=36$ MHz, $g_{2}/2\pi=36.6$ MHz (corresponding to $G=\frac{g_{1}g_{2}}{\Delta}=2\pi \times 4.4$ MHz), and $\Omega_{1}=10 \Omega_{2}=10^2G$. We assume that the qubit is initially in the excited state $|e\rangle$ and the magnon mode is in the vacuum state, which is the case of low bath temperature, e.g., of tens of mK.  Clearly, magnon squeezed states can be achieved and the two results (using the Hamiltonians~\eqref{e3} and \eqref{e9}) agree well with each other, indicating that our derived effective Hamiltonian is a very good approximation.  In Fig.~\ref{fig2}(b), a smaller value of $\Omega_{2}=5G$ is used, which just satisfies the condition $|\Omega_{2}|\gg\frac{G}{2}$ for deriving the Hamiltonian \eqref{e9}. The deviation of the two curves becomes larger especially for longer evolution time. Nevertheless, the effective Hamiltonian is still a good approximation once the conditions listed in Sec.~~\ref{model} are fulfilled.

\begin{figure}[t]
\centering
\hskip-0.3cm\includegraphics[width=\linewidth]{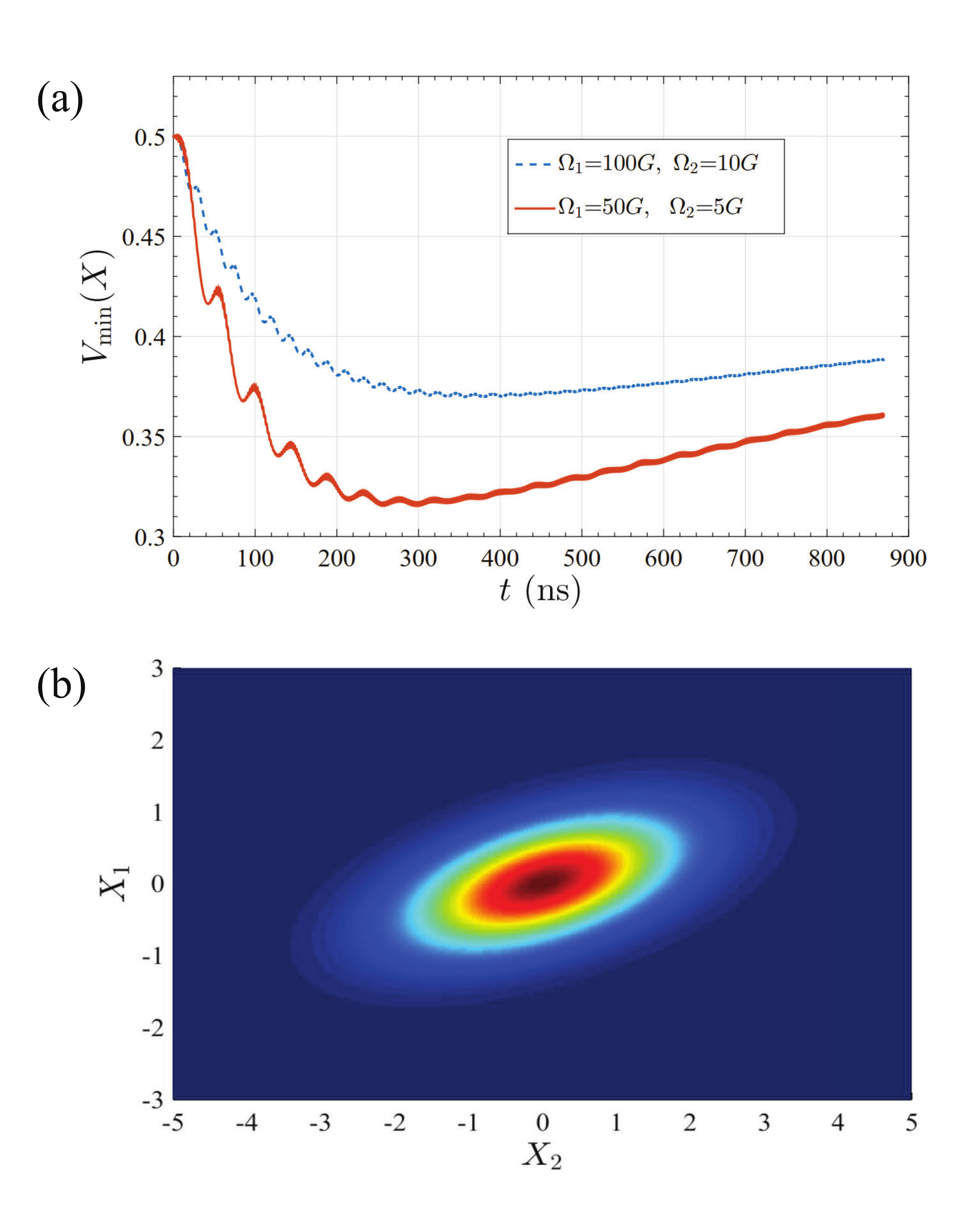}
\caption{(a) Minimum variance $V_{\rm min}(X)$ of the magnon quadrature versus $t$ with $\kappa/2\pi=1$ MHz and $\gamma/2\pi=20$ kHz for $\Omega_{1}=10 \Omega_{2} =10^2G$ (blue dashed line) and for $\Omega_{1}=10 \Omega_{2}=50G$ (orange solid line). (b) Wigner function of the magnon mode corresponding to the orange line in (a) at $t=300$ ns. The other parameters are the same as in Fig.~\ref{fig2}.}
\label{fig3}
\end{figure}

Figure~\ref{fig2} is obtained without considering any dissipation of the system. Therefore, the variance $V_{\rm min}(X) \to 0$ when $t \to \infty$. In what follows, we analyze the effect of the magnon and qubit dissipations on the degree of the squeezing. We adopt the Lindblad master equation \cite{79,Lindblad}
\begin{eqnarray}\label{e11}
\frac{d}{dt}\rho&=&-i[H, \rho]+\kappa(\bar{n}_{m}+1)\mathcal{L}_{m}\rho+\kappa \bar{n}_{m}\mathcal{L}_{m^{\dagger}}\rho\cr
&&+\gamma(\bar{n}_{q}+1)\mathcal{L}_{\sigma^{-}}\rho+\gamma \bar{n}_{q}\mathcal{L}_{\sigma^{+}}\rho,
\end{eqnarray}
where
\begin{eqnarray}\label{e12}
\mathcal{L}_{o}\rho=(o\rho o^{\dagger}-\frac{1}{2}o^{\dagger}o\rho-\frac{1}{2}\rho o^{\dagger}o)
\end{eqnarray}
represents the Lindblad term for an arbitrary operator $o$ $(o=m,m^{\dagger},\sigma^{-},\sigma^{+})$. $\kappa$ ($\gamma$) is the dissipation rate of the magnon mode (the qubit), and $\bar{n}_{m}$ ($\bar{n}_{q}$) is the mean thermal occupation number, and $\bar{n}_{j} \simeq [{\rm exp}(\hbar\omega/k_B T)-1]^{-1}$ ($j=m,q$) with $T$ being the bath temperature.  In Fig.~\ref{fig3}(a), $V_{\rm min}(X)$ is plotted with the dissipation rates $\kappa/2\pi=1$ MHz and $\gamma/2\pi=20$ kHz and at temperature $T=10$ mK~\cite{XuDa} for two sets of drive conditions, which correspond to those used in Figs.~\ref{fig2}(a) and \ref{fig2}(b), respectively.   Compared with the no-dissipation case of Fig.~\ref{fig2}, it is evident that the dissipations can significantly reduce the degree of the squeezing. Moreover, there is an optimal time for achieving the maximum squeezing, after which more noises enter the system through the dissipation channels and degrade the squeezing. To vividly show the magnon squeezing, we plot the Wigner function of the magnon mode in Fig.~\ref{fig3}(b), corresponding to the point at $t=300$ ns in the orange curve of Fig.~\ref{fig3}(a) and the minimum variance of $0.31$.

\begin{figure}[t]
\centering
\includegraphics[width=\linewidth]{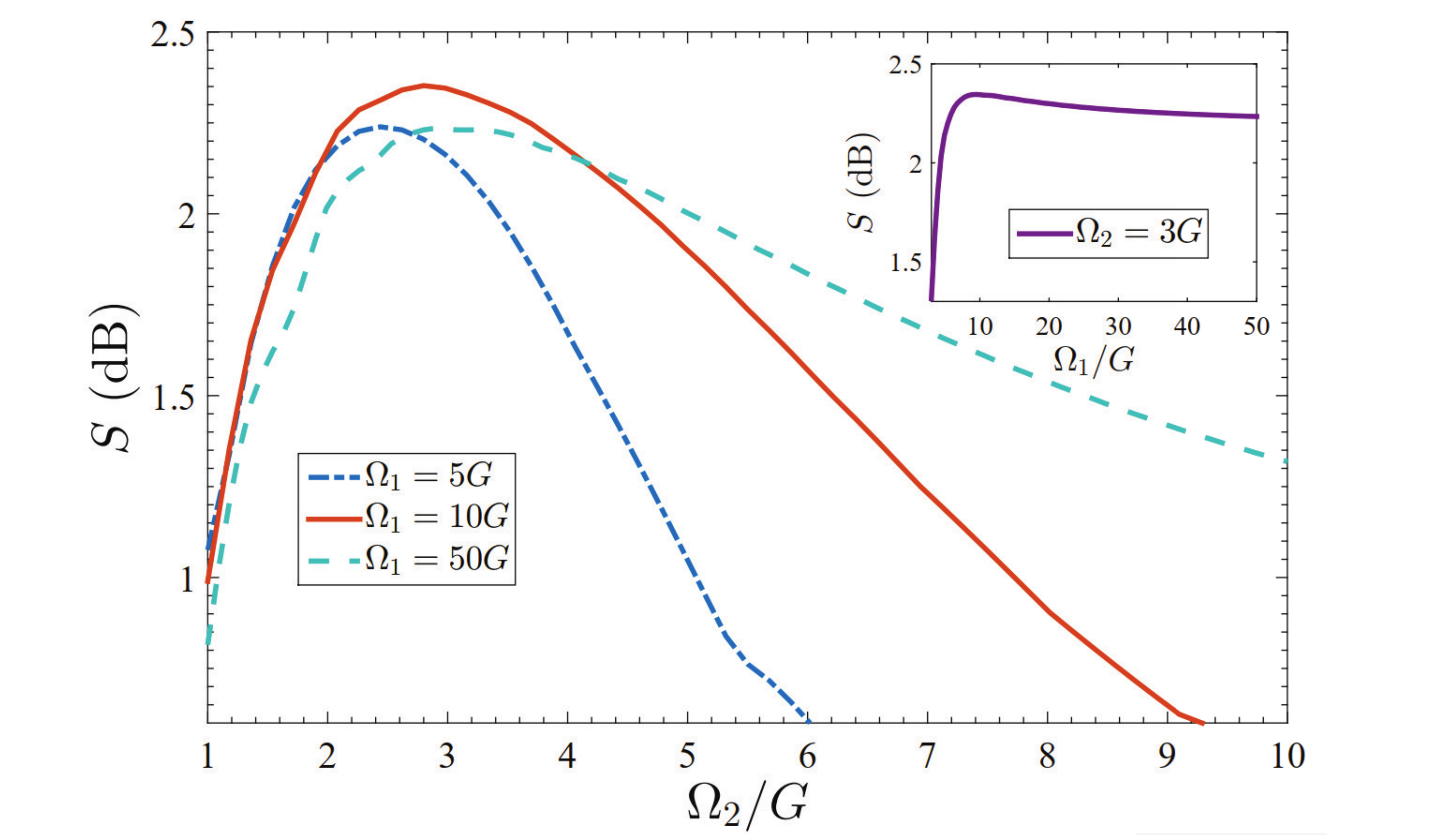}
\caption{The degree of squeezing S (dB) versus driving strength $\Omega_{2}$ for different values of $\Omega_{1}$. The inset shows S (dB) versus $\Omega_{1}$ for a fixed $\Omega_{2}=3G$. The other parameters are the same as in Fig.~\ref{fig3}. }
\label{f4}
\end{figure}

We now analyze the optimal drive conditions for obtaining the magnon squeezing. Summarizing the conditions used for deriving the desired parametric amplification Hamiltonian \eqref{e9}, we have $|\delta_2| = 2 \Omega_1 \gg \frac{\Omega_2}{2} \gg \frac{G}{4}$. Once the frequency of the second drive is determined (i.e., $\delta_2$ and $\Omega_1=\frac{|\delta_2|}{2}$ are fixed), it puts an upper limit on the driving strength $\Omega_2$ to get the optimal squeezing. A smaller $\Omega_2$ is preferred since the degree of squeezing is proportional to $\chi=\frac{G^2}{4\Omega_{2}}$.  However, $\Omega_2$ cannot be too small because of the lower limit of $\Omega_2 \gg \frac{G}{2}$. This further sets an upper limit on the maximum squeezing that can be achieved in our protocol since $\chi \ll \frac{G}{2}$. The presence of an optimal $\Omega_2$ is confirmed by Fig.~\ref{f4}, in which we have evaluated the degree of squeezing in units of dB, which is defined as $S=-10 {\rm log}_{10}[V_{\rm min}(X)/V_{\rm vac}(X)]$, where $V_{\rm vac}(X)=\frac{1}{2}$ corresponds to the vacuum fluctuation, and $V_{\rm min}(X)$ is obtained at the optimal time and at temperature $T=10$ mK. The dissipation rates considered in Fig.~\ref{f4} are the same as in Fig.~\ref{fig3}.

\begin{figure}[t]
\hskip-0.4cm \includegraphics[width=\linewidth]{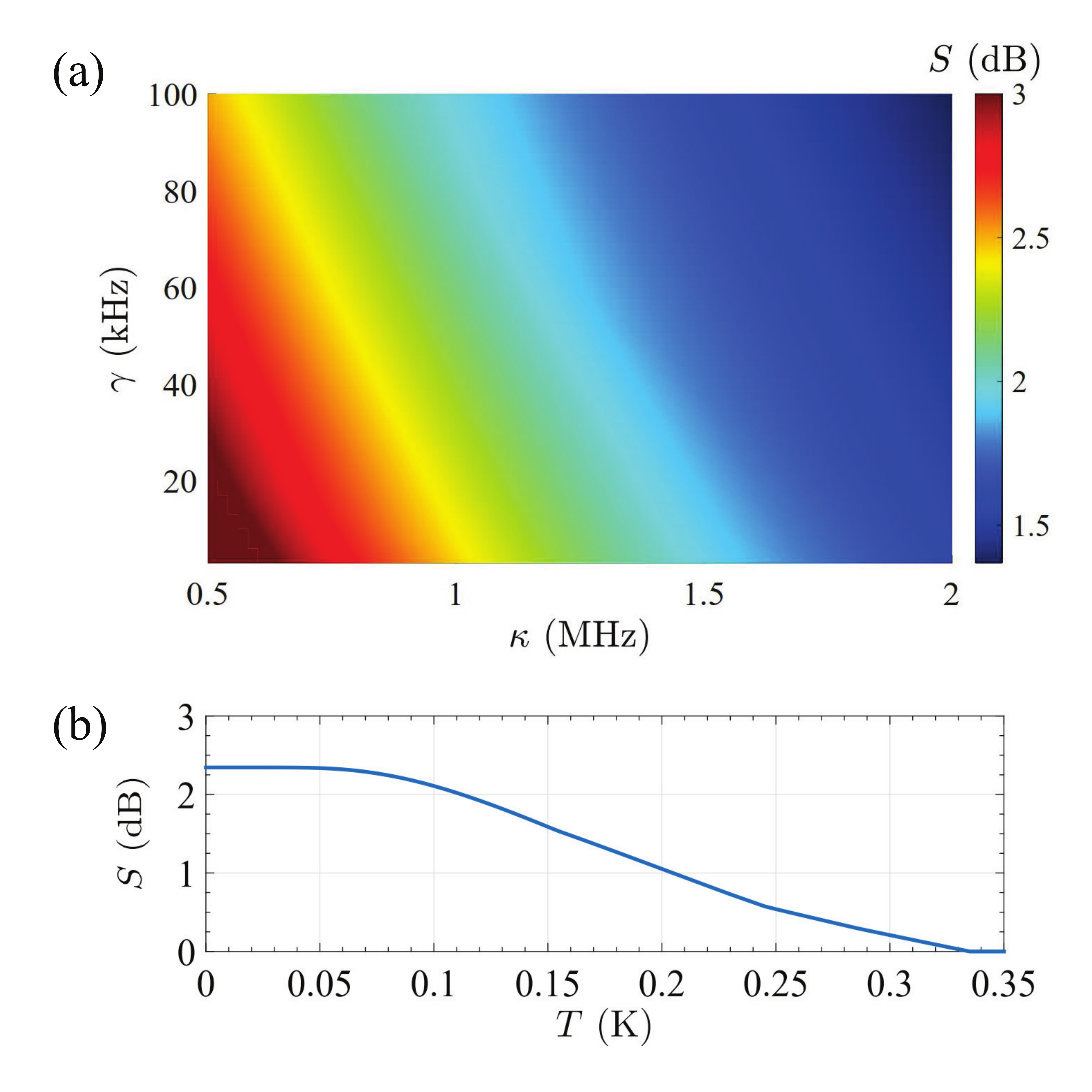}
\caption{The degree of squeezing $S$ (dB) versus (a) magnon and qubit dissipation rates $\kappa$ and $\gamma$ at $T=10$ mK;  (b) temperature $T$ for $\kappa/2\pi = 1$ MHz and $\gamma/2\pi = 20$ kHz. We take $\Omega_{1}=10G$ and $\Omega_{2}=3G$. The other parameters are the same as in Fig.~\ref{fig2}. }
\label{f5}
\end{figure}

In the inset of Fig.~\ref{f4}, we plot the degree of squeezing versus $\Omega_{1}$ for a fixed $\Omega_{2}=3G$. It shows that there is also an optimal driving strength $\Omega_{1}$. This is because, on the one hand, the driving strength must be strong enough to satisfy $\Omega_1 \gg \frac{\Omega_2}{4}$; while on the other hand, it cannot be too strong as a large $\Omega_{1}$ corresponds to a large detuning $|\delta_2|= 2 \Omega_1$, which reduces the drive efficiency associated with the second drive and thus the degree of squeezing. It should be noted that the drive frequencies $\omega_{1,2}$ are determined by $\omega_{Q}$ and $\Omega_{1}$, so according to Fig.~\ref{f4}, the optimal drive frequencies can also be determined.

The squeezing is robust against dissipations of the system and bath temperature, as shown in Fig.~\ref{f5}. We plot in Fig.~\ref{f5}(a) the degree of squeezing $S$ (dB) versus two dissipation rates $\kappa$ and $\gamma$ at low temperature $T=10$ mK.  Clearly, the squeezing is present for a wide range of both $\kappa$ and $\gamma$.  In Fig.~\ref{f5}(b), we plot $S$ versus $T$ for $\kappa/2\pi=1$ MHz and $\gamma/2\pi=20$ kHz~\cite{XuDa}. It shows that the squeezing is still present for the temperature up to $\sim330$ mK.

\section{Conclusions}\label{conc}

We present a scheme for preparing magnon squeezed states in a hybrid cavity-magnon-qubit system. The qubit is simultaneously driven by two microwave fields.  By properly selecting the drive frequencies and strengths, an effective parametric amplification Hamiltonian is obtained for the magnon mode, which yields magnon quadrature squeezing. We provide the optimal drive conditions and analyze the validity of the model. The magnon squeezing is robust against dissipations and bath temperature, and the numerical results indicate that moderate squeezing can be achieved using fully realistic parameters from recent experiments \cite{23,25,26,XuDa}. The squeezed state, with the noise below vacuum fluctuation, is of a magnon mode consisting of more than $10^{18}$ spins for a 1-mm-diameter-YIG sphere and thus represents a macroscopic quantum state. The work may find potential applications in the study of macroscopic quantum phenomena, as well as in high-precision measurements based on magnons.

\section*{Acknowledgments}
We thank Dr. Da Xu for useful discussion on the experimental feasibility.  This work has been supported by National Key Research and Development Program of China (Grant No. 2022YFA1405200) and National Natural Science Foundation of China (Grant Nos. 12274274, 12174140, and 92265202).

\end{document}